# Free-electron radiation engineering via structured environments

Hao Hu[1], Xiao Lin[2,3], and Yu Luo[1,*]

[1]*School of Electrical and Electronic Engineering, Nanyang Technological University, 50 Nanyang Avenue, Singapore 639798, Singapore.*
[2]*Interdisciplinary Center for Quantum Information, State Key Laboratory of Modern Optical Instrumentation, College of Information Science and Electronic Engineering, Zhejiang University, Hangzhou 310027, China.*
[3]*International Joint Innovation Center, ZJU-UIUC Institute, Zhejiang University, Haining 314400, China.*
[*]*E-mail: luoyu@ntu.edu.sg (Y. Luo)*

**ABSTRACT**

Free-electron radiation results from the interaction between swift electrons and the local electromagnetic environment. Recent advances in material technologies provide powerful tools to control light emission from free electrons and may facilitate many intriguing applications of free-electron radiation in particle detections, lasers, quantum information processing, etc. Here, we provide a brief overview on the recent theoretical developments and experimental observations of spontaneous free-electron radiation in various structured environments, including two-dimensional materials, metasurfaces, metamaterials, and photonic crystals. We also report on research progresses on the stimulated free-electron radiation that results from the interaction between free electrons and photonic quasi-particles induced by the external field. Moreover, we provide an outlook of potential research directions for this vigorous realm of free-electron radiation.

# INTRODUCTION

Free-electron radiation refers to the physical process that moving charged particles (e.g. free electrons) emit light into the surrounding environment. One prototypical free-electron radiation is Cherenkov radiation, which was experimentally observed by P. Cherenkov in 1934 [1] and later theoretically explained by I. Tamm and I. Frank in Ref [2]. According to their theory, the uniformly moving charged particle can produce Cherenkov radiation when the particle velocity exceeds a threshold velocity, i.e., the phase velocity of light in the surrounding medium. Cherenkov radiation is of key importance in the community of particle physics and photonics owing to its unique features such as high emission directionality, continuous spectrum, etc. As a concrete example, Cherenkov radiation has indispensable applications in high-energy particle detection [3-5], because of the strong dependence of Cherenkov emission angle on the particle velocity.

Subsequent theoretical and experimental developments have revealed that emission behaviors (e.g., threshold velocity, emission angle, polarization, and intensity) of free electrons can be tailored. First of all, emission behaviors of free electrons highly depend on the particle trajectory. For instance, when the free electron moves in a circular trajectory, the electron emits harmonics of its cyclotron frequency. Such a radiation phenomenon is known as synchrotron radiation (for relativistic particle velocity) or cyclotron radiation (for nonrelativistic particle velocity) [6]. More importantly, the surrounding environment also strongly influences emission behaviors, leading to many fundamentally new classifications of free-electron radiation, including transition radiation, Smith-Purcell radiation, and bremsstrahlung radiation. To be specific, transition radiation refers to the scattering radiation when a charged particle passes through a spatial (or temporal) boundary of two different media [7, 8]; Smith-Purcell radiation corresponds to the diffraction radiation when a free electron moves close to a periodic grating [9]; Bremsstralung radiation is the braking radiation when the swift charged particle is decelerated [10]. Enabled by their tailorable emission properties, free-electron radiations find broad applications in the field of particle physics, integrated photonics, microscopy, medical imaging and therapy, etc.

On the other hand, structured environments have gained increasing attentions thanks to the development of modern material technologies. Such structured environments include but not limited to two-dimensional materials, metasurfaces, metamaterials, and photonic crystals. Two-dimensional materials are substances with a thickness down to atomic scale [11]. Electrons/phonons in these materials are free to move in a two-dimensional plane, leading to the formation of surface plasmon/phonon polaritons. Highly confined surface modes in two-dimensional materials have important applications in nanoscale manipulation of electromagnetic waves. Metamaterials consists of a densely arranged array of optical scatters in the subwavelength scale [12]. As a key advantage, the effective macroscopic parameters of metamaterials can go beyond those of natural existing materials. The prominent examples of metamaterials include negative-index, zero-index and hyperbolic metamaterials. The continuously emerging metamaterials provide new paradigms for the photon manipulation. Photonic crystals (i.e., photonic analogues of atomic lattices) are made of periodically arranged optical scatters with their lattice constant comparable to the wavelength [13]. Intriguingly, photonic crystals can reflect and trap light in a designed way that is prohibited for their constituents. Recent advances reveal the existence of topological phases in photonic crystals, and many photonic topological insulators have been discovered, including photonic Chern insulators, photonic Weyl semimetals, etc [14]. These findings pave the way to novel photonic applications that are robust against external perturbations.

The wide variety of structured materials provides powerful tools to control emission behaviors of free-electron radiation that leads to many significant findings [15-20]. Simply within the topic of Cherenkov radiation, the relevant studies have continuously demonstrated new phenomena include non-divergent [21], threshold-free [22] and reversed Cherenkov radiation [23]. In addition, recent theoretical and experiment work have shown the important role of quantum nature on free-electron radiation in structured environments while these quantum effects are conventionally negligible in natural existing materials [24]. Such a quantum interaction is even more significant in the presence of external fields, giving birth to the important applications such as the photon-induced near-field electron microscopy [25].

The engineered free-electron radiation in structured environments thus shows a promising future for the next generation of particle detectors, free-electron lasers, particle accelerators, electron microscope, etc.

**SYSTEMS FOR FREE-ELECTRON RADIATION ENGINEERING**

*Two-dimensional materials*

The community of nanophotonics has drawn significant research attention on two-dimensional materials in past few years owing to their ability to control the light-mater interaction in a squeezed dimension [26]. The highly desired free-electron radiation source at the nanoscale naturally motivates the study of free-electron radiation in the platform of two-dimensional materials.

One prominent example of two-dimensional materials is graphene that supports surface plasmon polaritons resulting from the oscillation between light and electrons in materials [27-30]. Compared to other plasmonic materials such as noble metals, graphene has shown so many intriguing properties to manipulate free-electron radiation [31]: first, the high degree of confinement and low level of loss in the terahertz to mid-infrared frequency range makes graphene a potential platform for achieving on-demand free-electron radiation sources; second, the extremely small phase velocity of graphene plasmon (down to 1/300 of light speed in the vacuum) relaxes the condition of free-electron radiation by significantly lowering down the threshold velocity; moreover, the flexible tunability of chemical potential of graphene can enable novel active optical devices.

These unique features facilitate the research of free-electron radiation in the platform of graphene [32-36]. A recent theoretical work has revealed the dynamical generation process of surface plasmons by a free electron perpendicularly penetrating through the graphene as shown in Figure 1(a) [37, 38].The calculation results have shown that the swift electron dissipates a large part of energy before the graphene plasmons are excited, and deepens our understanding on the dynamic process of electron energy loss. Another relevant work bridges the gap between the free-electron radiation and nonlinear optics in Figure 1(b) [39]. Thanks to the high tunability of graphene plasmons in graphene nano-islands, the condition of emitting nonlinear signals from free electrons can be largely relaxed. As a result, the low-energy electron can trigger strong nonlinear response in the graphene, in sharp contrast to the case of metallic structures where high-energy electrons are required.

Graphene is also an exciting platform to access the quantum regime of free-electron radiation. Cherenkov radiation from charge carriers in graphene is widely believed to be prohibited because the velocity of carriers limited by the Fermi velocity is impossible to exceed the classical Cherenkov threshold. However, recent study has shown that the actual Cherenkov threshold can be below the Fermi velocity, if the quantum recoil of charge carrier comes into play [40]. Thus, charge carriers in graphene can indeed enable efficient two-dimensional Cherenkov radiation [Figure 1(c)]. Another exciting and relevant study is related to the Compton scattering of the free electrons interacting with graphene plasmons [Figure 1(d)] [41]. Relying on the large momentum of graphene plasmons, the resulted emission from the free electron is featured by extremely high frequency and directional nature. These fundamental research holds promises for achieving highly efficient on-chip sources applicable to terahertz and X-ray frequencies.

Since the advance of graphene, the interest in the atomically layered materials has grown exponentially. More recently, two-dimensional polar van der Waals materials have been realized [42]. These materials support phonon polaritons that originates from the oscillation between the light and lattice vibrations. The most widely studied two-dimensional polar van der Waals material is hexagonal boron nitride (hBN), that can support phonon polaritons with strong field confinement, long lifetime and extreme polariton anisotropy in mid-infrared frequencies. These striking properties stimulate the recent research on Cherenkov radiation of phonon polaritons in hBN [43-45]. The newly revealed Cherenkov radiation opens up an opportunity to realize the tunable free-electron radiation source in the mid-infrared frequency [Figure 1(e,f)] [46].

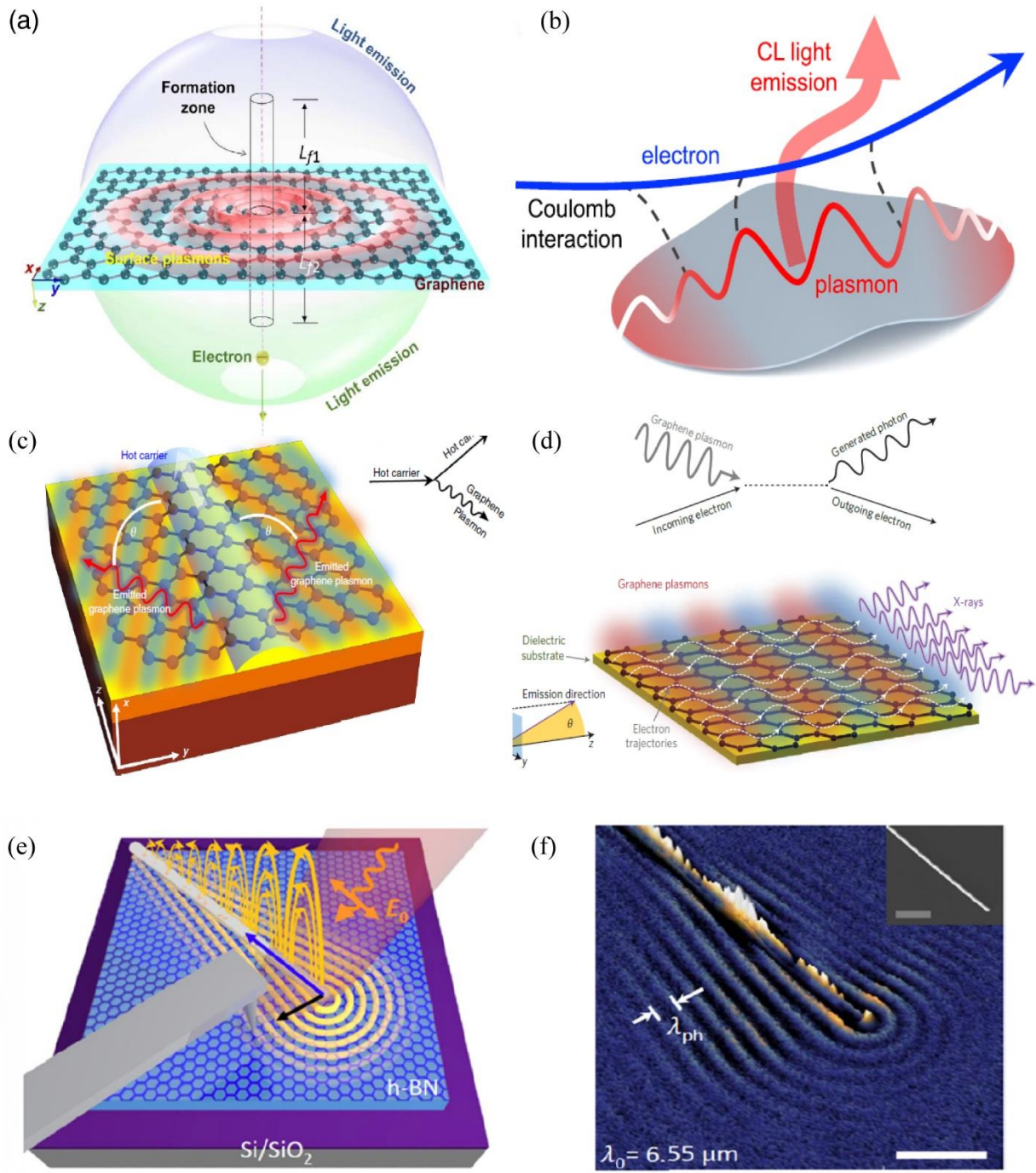

**Figure 1.** Free-electron radiation in two-dimensional materials. (a) Transition radiation from a graphene sheet. (b) Nonlinear interaction between the free electron and graphene nano-island. (c) Cherenkov radiation from hot carriers in the graphene. (d) Compton scattering from the free electron interacting with graphene plasmons. In (c,d), the Feynman diagrams are shown in the corresponding inset. (e-f) Cherenkov radiation in silver nanowire/hBN heterostructures. (e) Schematic of experimental realization. The silver nanowire is illuminated by a laser beam with the electric field parallel to the nanowire. The excited plasmon propagates along the nanowire. The propagating metallic plasmon generates Cherenkov radiation of hBN's phonon polaritons. (f) Experimental image of Cherenkov radiation of phonon polaritons in the silver nanowire/hBN heterostructures. (a) Reproduced with permission [37]. Copyright 2017, American Association for the Advancement of Science. (b) Reproduced with permission [39]. Copyright 2020, American Chemical Society. (c) Reproduced with permission [40]. Copyright 2016, Nature Publishing Group. (d) Reproduced with permission [41]. Copyright 2016, Nature Publishing Group. (e,f) Reproduced with permission [46]. Copyright 2020, American Chemical Society.

*Metamaterials*

The interaction between the free electron and metamaterials has attracted wide attentions in the last two decades. A large class of novel free-electron radiation enabled by metamaterials have been revealed. For example, non-divergent Cherenkov radiation is discovered in the nanowire metamaterial [Figure 2(a)] [21]. By engineering flat eigenfrequency contour for nanowire metamaterials, the field amplitude of Cherenkov radiation remains unchanged with the increasing distance from the electron trajectory along the radiation path [Figure 2(b)]. Another remarkable emission behavior enabled by metamaterials is the reversed Cherenkov radiation. This phenomenon has been theoretically predicted and experimentally observed on the left-handed metamaterials [23, 47-49], whose permittivity and permeability are designed to be simultaneously negative [Figure 2(c)]. Due to the negative refractive index of structured environments, the traditional forward Cherenkov cone is reversed, counter-intuitively leading to the backward radiation [see the measured power in port 1&2 in Figure 2(d)]. In addition, the transformation optics offers a more general method to design metamaterials that can manipulate the free-electron radiation in a flexible way. As a demonstration, free-electron radiation is studied in the platform of invisible cloak designed by transformation optics [50]. When the free electron travels through the invisible cloak, the emission behavior remains unperturbed even in the presence of objects inside the cloak [Figure 2e]. More recently, Cherenkov radiation in the hyperbolic metamaterials has excited broad research interest. Ref. [22] has mentioned that Cherenkov radiation in the hyperbolic metamaterials is threshold-free, owing to the their extremely large photonic density of states [Figure 2(g)] [22]. In their experiment, the minimum measured threshold velocity is down to 0.25 eV [22] . However, the further reduction of the electron velocity increases the difficulty for the measurement of Cherenkov signal. Recently, Ref. [51] found the threshold of Cherenkov radiation in hyperbolic metamaterials is actually always nonzero by considering the spatial nonlocality. To be specific, the nonzero Cherenkov threshold [51] is determined by the spatial dispersion from the finite period and nonlocal electron screening in metals [Figure 2(h)]. These findings are of fundamental importance to develop novel free-electron radiation sources with precisely controllable emission angles, threshold velocities, etc.

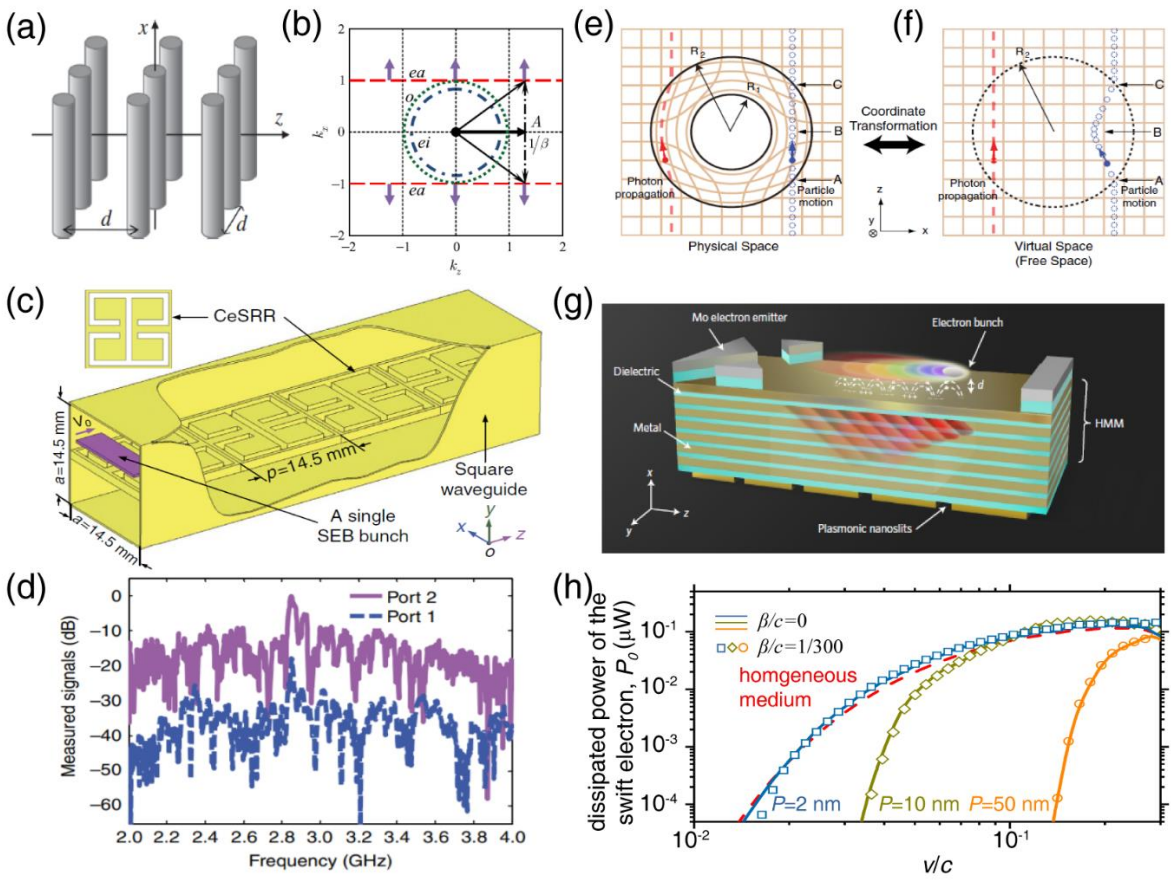

**Figure 2.** Free-electron radiation in metamaterials. (a,b) Cherenkov radiation in a nanowire metamaterial. (a) Schematic of nanowire metamaterial. The period of nanowire metamaterial is *d*. The free electron transverses the nanowire metamaterial along the *z*-direction. (b) Isofrequency contour for the ordinary wave (dotted line), "isotropic" extraordinary wave (dashdotted line) and "anisotropic" extraordinary wave (dashed line). The directions of phase velocity and group velocity are indicated by the long and short arrows, respectively. (c-d) Cherenkov radiation in a left-handed metamaterial. (c) Experimental setup. An electron bunch is moving on the top of left-handed metamaterial inside the square waveguide. (d) Measured power spectral densities. Port 1 measures the intensity of forward emission while port 2 measures the intensity of backward emission. (e,f) Cherenkov radiation in an invisible cloak. The cloak region is indicated by the dashed circle. Photon propagation path (denoted as red dashed line) and free particle trajectory (denoted as bule circles) are highlighted in both the physical space (e) and the virtual space (f). (g-h) Cherenkov radiation in a hyperbolic metamaterial, which is constructed by a metallo-dielectric layered structure. (g) Experimental setup. An electron bunch is moving on the top of metallo-dielectric layered structure. (h) The total radiation power of free electron in hyperbolic metamaterial. Here, *P* is the structure period describing the degree of spatial nonlocality from the finite period; *β* is the nonlocal parameter describing the degree of nonlocal electron screening in metals. All the nonlocal effects are neglected when the layered structure is replaced by the homogeneous medium (as indicated by the red dashed line). (a,b) Reproduced with permission [21]. Copyright 2012, American Physical Society. (c,d) Reproduced with permission [48]. Copyright 2017, Nature Publishing Group. (e,f) Reproduced with permission [50]. Copyright 2009, American Physical Society. (g) Reproduced with permission [22]. Copyright 2017, Nature Publishing Group. (h) Reproduced with permission [51]. Copyright 2020, Wiley VCH.

***Photonic crystals***

Photonic crystals provide an additional route to control free-electron radiation. Similar to that of metamaterials, the isofrequency contour of photonic crystals can be flexibly engineered for photon manipulation. Early in 21$^{st}$ century, Ref. [52] has demonstrated how the engineered isofrequency contour of photonic crystal can flexibly impact emission behaviors (i.e., emission angles and threshold velocity) of Cherenkov radiation in Figure 3(a,b) [52]. On the other hand, photonic crystal has a photonic bandgap resulting from the diffraction and destructive interference. This is different from metamaterials, whose diffraction behaviors are forbidden. Such a unique feature has been used to design tunable and efficient free-electron radiation sources in Figure 3(c) [53]. To be specific, when the free electron transverses the photonic crystal, photon emission from the free electron is enhanced when the evanescent wave of electron couples to the eigenmodes of photonic crystal outside the photonic bandgap, whereas the photon emission is prohibited within the photonic bandgap. Via judicious structural design, the modified dispersion band can lead to the frequency shift of designed radiation sources.

Frontier research in free-electron radiation has explored the interaction between free electrons and novel optical quasi-particles in photonic crystals. One special quasi-particle in photonic crystal is the bound state in the continuum (BIC) that behaves localized but can couple to extended waves. Owing to its negligible radiation loss, BICs are featured by superhigh quality factor (Q factor) close to the infinity [see inset in Figure 4(e)] and this feature makes them particularly important for laser applications. Recent works have demonstrated the free-electron radiation coupled to the BICs in the photonic grating [Figure 4(d-g)] [54, 55]. The calculation results have shown that the free electron can efficiently excite the BICs at a particular velocity when the electron wavevector is infinitely close to the wavevectors of BICs. The high-Q free-electron radiation enabled by BICs has potential applications in free-electron laser. In parallel to the development of BICs, photonic topological states have been widely studied recently. Photonic topological states propagate at the interface of topological photonic crystals as a manifestation of bulk-edge correspondence. The properties of topological states are protected by a particular geometric symmetry, and these modes are robust against certain disorders. Free-electron radiation coupled to topological edge states has been recently revealed in Ref. [56] [Figure 4(h-k)]. This work has designed two quantum spin-hall photonic topological insulators with opposite topological invariants (i.e., spin Chern numbers) [Figure 4(j)]. Two topological edge states emerge at the domain wall between two types of topological photonic crystals. The propagation direction of topological edge states is stringently locked to their propagation direction: spin-up (spin-down) leads to the forward (backward) propagation. A free electron traversing the domain wall at the angle of 60° relative to the interface successfully excites the spin-up topological edge state that propagates toward right direction. Reversing the top and bottom photonic topological insulator flips the spin of excited topological edge states, and thus leads to the wave propagation at an opposite direction. Owing to the topologically protected transportation of edge states by the domain wall, the measured emission intensity in the terminal of domain wall is significantly larger than that in the bulk [Figure 5(k)]. The revealed transition radiation coupled to topological states opens a new door to achieve the novel free-electron radiation source with topologically enabled robustness.

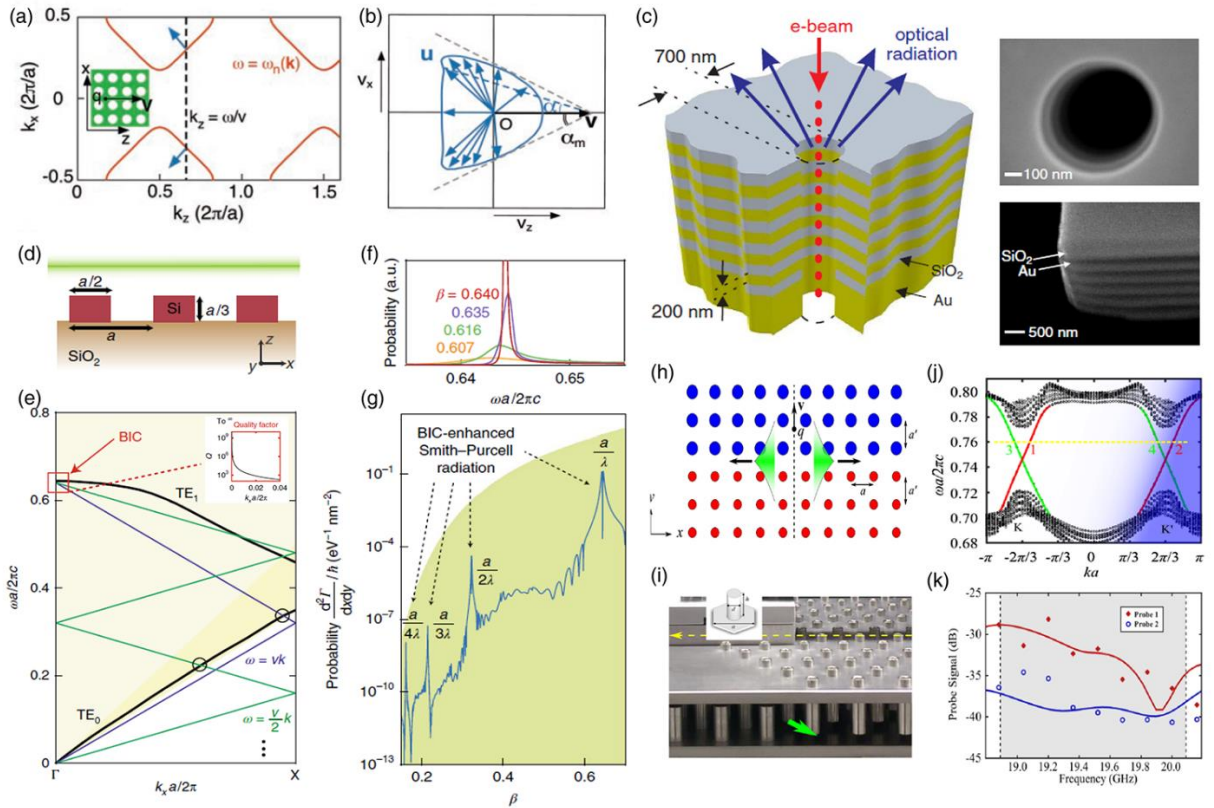

**Figure 3.** Free-electron radiation in photonic crystals. (a,b) Controlling Cherenkov angles with the two-dimensional all-dielectric photonic crystal. (a) Schematic of photonic crystal and its isofrequency contour. (b) Contours of group velocities in the photonic crystal. $\alpha$ is the angle of Cherenkov cone when the particle velocity is $v$, and the group velocity is $u$. (c) Tunable free-electron radiation source enabled by the one-dimensional photonic crystal, which is a periodic stacking of metal and dielectric slabs. (d-g) Free-electron radiation coupled to the BICs. (d) Structural schematic. The free electron moves close to and on the top of an all-dielectric photonic grating. (e) Band structure. The dispersion band of the photonic grating in (d) is marked in black whereas the electron line is marked in green. The inset shows the $Q$ factor of modes around the BIC. (f) Spectra of free-electron radiation at different electron velocities. (g) Emission intensity versus the electron velocity. (h-k) Free-electron radiation coupled to topological edge states. (h) Structural schematic. The blue and red cylinder arrays denote two topologically distinct domain. (i) Demonstration of the experimentally designed photonic topological insulator. (j) Band structure. The red (blue) curve denotes the spin-up (spin-down) topological edge state whereas the black curve denotes bulk states. (k) Measured emission spectrum from the free electron. Probe 1 (probe 2) is close to (far from) the end of domain wall. (a,b) Reproduced with permission [52]. Copyright 2003, American Association for the Advancement of Science. (c) Reproduced with permission [53]. Copyright 2009, American Physical Society. (d-g) Reproduced with permission [55]. Copyright 2018, Nature Publishing Group. (h-k) Reproduced with permission [56]. Copyright 2019, American Physical Society.

*External fields*

Free-electron radiation without external fields belongs to the spontaneous emission, while free-electron radiation in the presence of external fields correspond to the stimulated process. The theoretical and experimental development of stimulated free-electron radiation leads to the important applications in the photon-induced near-field electron microscopy (PINEM). Recent advances have shown the powerful capability of PINEM to characterize the photonic quasiparticles (serving as external fields) including surface plasmon polariton [Figure 4(a,b)] [57], phonon polariton [Figure (f)] [58], photonic crystal modes [Figure (c-e)] [59], etc. Figure 4(a) demonstrates the schematic to image the surface plasmons with PINEM [57]. Two light pulses are used in the experimental setup: an optical pulse irradiates the sample to excite the intended photonic quasi-particle (i.e., surface plasmons) whereas an ultraviolet pulse triggers the electron gun to launch the electron beam. The interaction between the free electron and surface plasmons results in their inelastic exchange of energy quanta, and thus the electron energy gain or loss. Via measuring the energy gain or loss with the electron energy spectrometer, one can learn the properties of studied mode such as amplitude, phase, and polarizations. More excitingly, PINEM can enable ultrafast time resolution of corresponding modes by controlling the time delay between two light pulses [see the example of surface plasmons in Figure 4(b)].

Besides, the stimulated free-electron radiation provides an excellent platform to shape the electron wave function. A recent work has shown the quantized electron energy comb when a free electron interacts with the evanescent wave (induced by the total reflection of optical pulse) [60]. As shown in Figure 4(g), the work has found that when the electron wavevector matches with the wavevector of evanescent wave, the light-electron interaction induces a resonant exchange of hundreds of photon quanta with the single electron. As a result, the electron wavefunction evolves into a quantized energy comb extending over hundreds of electron volts. On the other hand, the stimulated free-electron radiation can enable the generation and control of electron beam vortex as shown in Figure 4(h,i) [61]. In the experiment, two elliptically polarized optical pulses are used to excite interfering chiral plasmons on the plasmonic material [Figure 4(h)]. After the momentum transfer between the electron beam and interfering chiral plasmons, the electron beam carries nonzero vortex. Interestingly, the intensity and helicity of electron wave function can be dynamically tuned via controlling the time delay between two optical pulses and the helicity of incidences [Figure 4(i)].

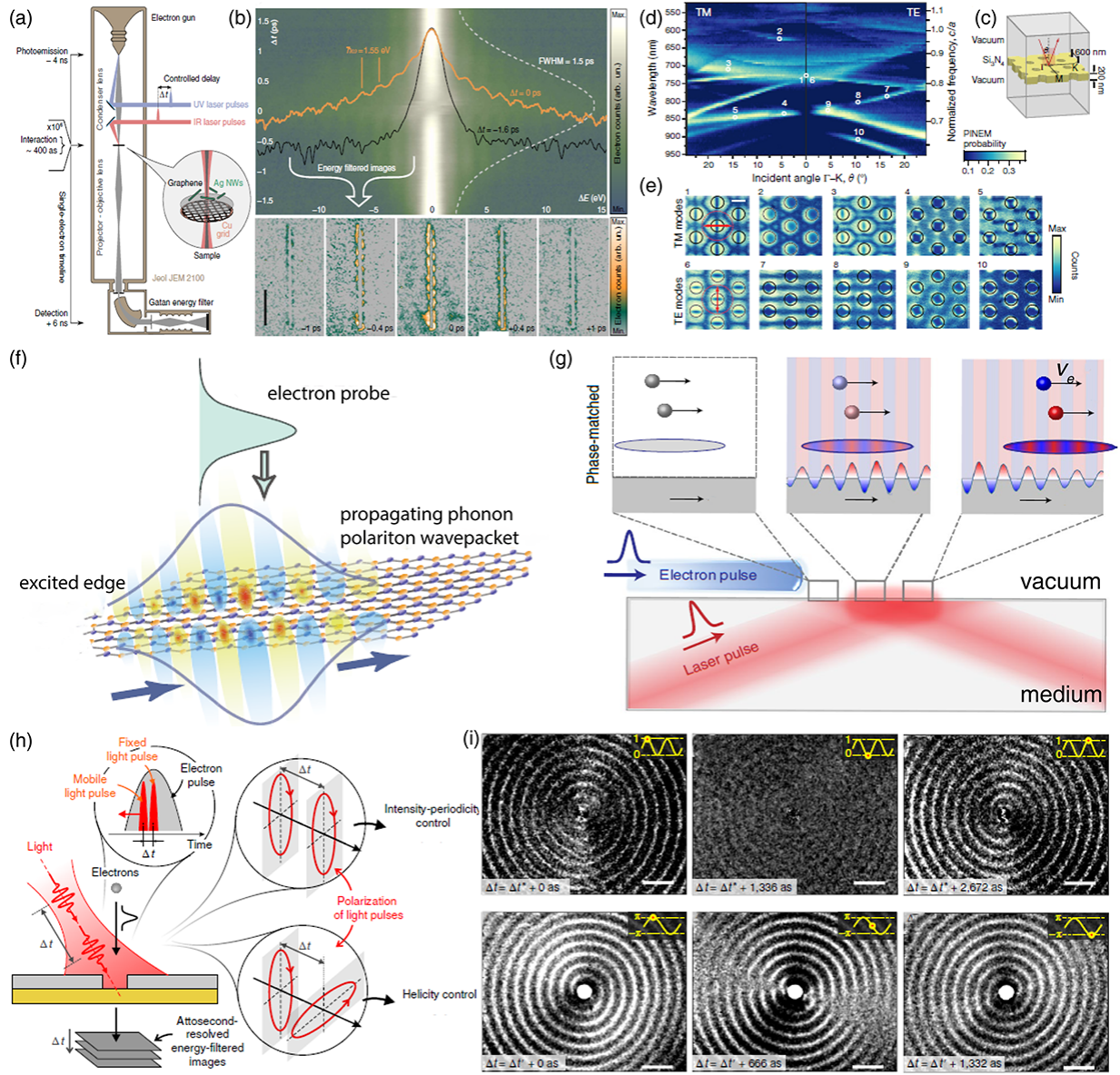

**Figure 4.** Free-electron radiation in external fields. (a-b) Imaging surface plasmon polaritons on a single nanowire by PINEM. (a) Schematic demonstration of the experimental setup. (b) Top: electron energy loss spectrum versus time delay between the optical pulse and electron pulse; Bottom: the inelastically scattered electron at different time moment. (c-e) Imaging the cavity modes of photonic crystals with PINEM. (c) Structural demonstration of the photonic crystal. (d) Measured band structure of the photonic crystal. (e) Spatial maps of the inelastically scattered electron from the photonic crystal. (f) Imaging hBN's phonon polaritons by PINEM. (g) Quantized electron energy comb enabled by the interaction

between the free electron and evanescent waves. The phase of light fields and electron wavepacket are indicated by the alternating red and blue. In the phase-matched condition, each point in the electron wavefunction always interacts with the light field in a resonant behavior. (h,i) Electron vortex beam enabled by the interaction between the free electron and interfering chiral plasmons. (h) Schematic of generating and controlling electron vortex beam. (i) Spatial maps of the inelastically scattered electron for intensity-periodicity control (top) and helicity control (bottom). (a,b) Reproduced with permission [57]. Copyright 2015, Nature Publishing Group. (c-e) Reproduced with permission [59]. Copyright 2020, Nature Publishing Group. (f) Reproduced with permission [58]. Copyright 2021, American Association for the Advancement of Science. (g) Reproduced with permission [60, 61]. Copyright 2020, Nature Publishing Group. (h-i) Reproduced with permission [61]. Copyright 2019, Nature Publishing Group.

**OUTLOOK**

To conclude, we have thoroughly reviewed the theoretical and experimental development of modified free-electron radiation in structured environments. The demonstrated novel free-electron radiation could find applications in particle physics, nanophotonics, microscopy, astrophysics, etc. Even so, we highlight that many research directions related to free-electron radiation remain unexplored.

So far, most studies focus their discussion on the spatially modulated media to manipulate the free-electron radiation, while how the temporally modulated media (or more generally, spatiotemporally modulated media) impact the radiation behaviors has been rarely researched [62-64]. Spatiotemporally modulated media are fundamentally different from the spatial media. First, the system constructed of spatially modulated media preserves the energy conservation that fundamentally limits the maximum energy loss of free electron. However, such a limitation is absent if the media are modulated in time. For example, the temporal photonic crystal can enable the parametric amplification once the wavevector is within the inverted bandgap, providing a valuable approach to realize the free-electron radiation with growing energy. Second, the spatial media preserve time-reversal symmetry while the spatiotemporal media spontaneously break such a symmetry. Third, the modulation velocity of spatiotemporal media can be subluminal or superluminal while that of spatial media is zero. These difference between the spatiotemporal media and spatial media raises fundamental questions regarding free-electron radiation in spatiotemporal media: What is the ultimate limit of the energy loss of free electrons? How the broken time reversal symmetry and different modulation velocity in spatiotemporal media influence the free-electron radiation, or will we witness new phenomena?

Another major problem remains to be tackled is related to the forbidden transition of free electrons. The forbidden transition refers to the transitions fundamentally forbidden by negligible decay rates include multipolar transitions (e.g., electric quadrupole and magnetic dipole transition), spin-flip transitions, multi-quasi-particle emissions, etc. Recent work has shown that the forbidden transitions of a molecule or atom can be observed via shrinking the light toward atomic length scale since the transition probability scales with field confinement. However, the forbidden transition of free electrons has yet been explored, even in theory. Realizing forbidden transition of free electrons may produce important applications such as entangled free-electron radiation source for quantum communications.

In addition, the realization of on-chip particle detectors is highly desired. Recently, attempts have been made to design high-performance particle detectors with nanotechnologies. Transformation-optical metamaterials [Figure 5(a)] [65] and resonance transition radiation in photonic crystals [Figure 5(b-c)] [66] have been proposed to achieve highly sensitive particle detectors with miniaturized volume. However, the narrow bandwidth of these two types of particle detectors severely hinders their practical applications. To overcome this problem, additional two mechanisms have been proposed, i.e., Brewster Cherenkov radiation [67] and surface Dyakonov-Cherenkov radiation [68]. Brewster Cherenkov radiation refers to Cherenkov radiation on the Brewster photonic crystal [Figure 5(d)]. Relying on the broadband optical angular selectivity of Brewster photonic crystal, the designed particle detector can enable particle detection with high sensitivity and broad bandwidth. Surface Dyakonov-Cherenkov radiation refers to Cherenkov radiation coupled to Dyakonov surface wave propagating on the surface of transparent birefringent materials [Fig. 5(e-h)]. The highly directional nature of surface Dyakonov-Cherenkov radiation not only enables particle detection with broad bandwidth, but also offers a scheme to measure the particle direction. Although these findings are exhilarating, there is no such an experiment to demonstrate a truly nanophotonic particle detectors yet. Such an experiment will eventually inspire the future nanotechnology such as on-chip particle detections.

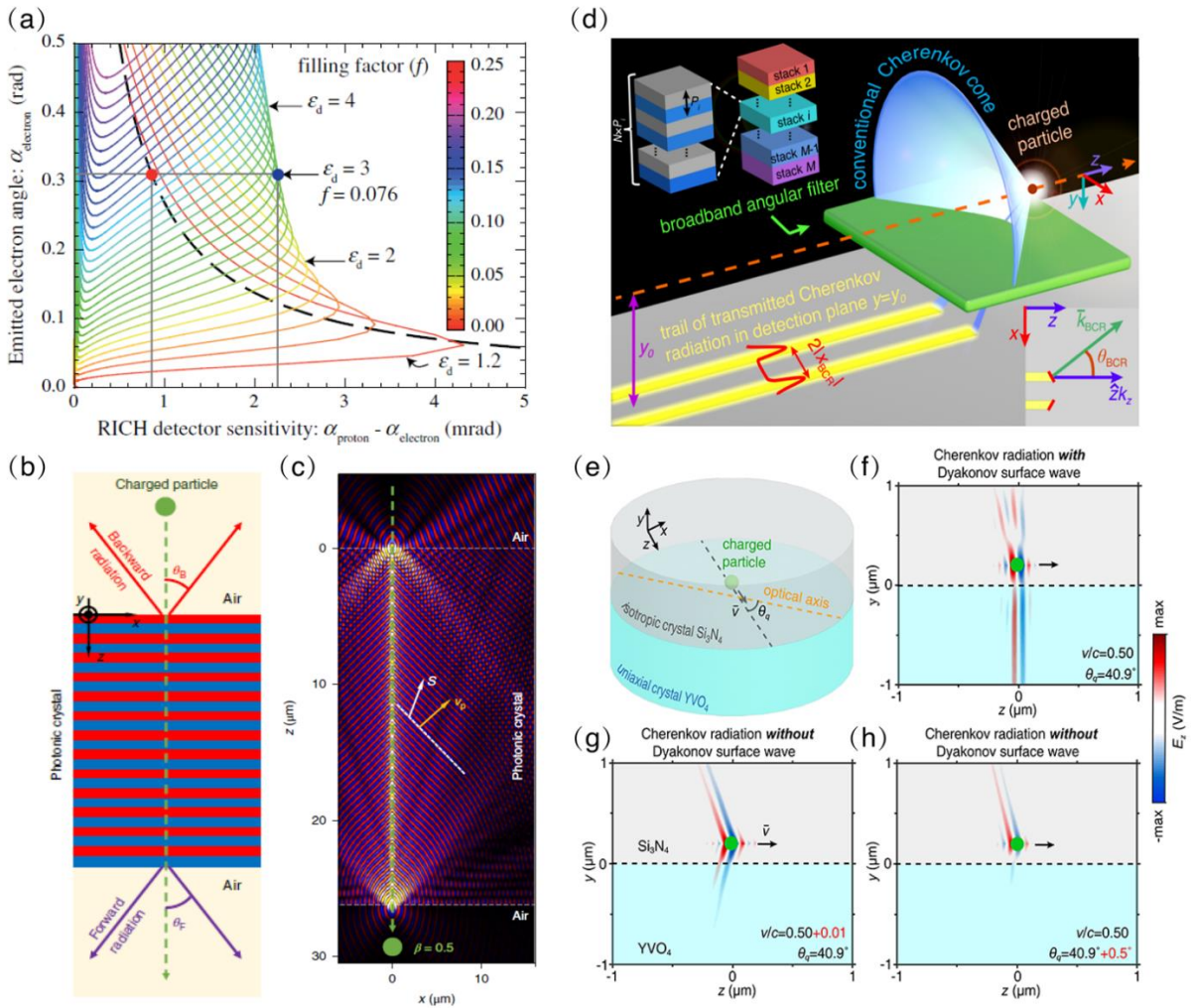

**Figure 5.** Examples of nanophotonic particle detectors. (a) Sensitivity of particle detectors enabled by the transformation-optical metamaterial. (b-d) Particle detectors based on the resonance transition radiation in photonic crystals. (b) Schematic demonstration of resonance transition radiation in photonic crystal. (c)

Field distribution of resonance transition radiation in the photonic crystal. (d) Particle detectors using Brewster Cherenkov radiation. Brewster photonic crystal is indicated as the green slab. The inset shows the details of Brewster photonic crystal. (e-h) Particle detectors using surface Dyakonov-Cherenkov radiation. (e) Structural demonstration. (f) Cherenkov radiation with Dyakonov surface waves. (g,h) Cherenkov radiation without Dyakonov surface waves. (a) Reproduced with permission [65]. Copyright 2014, American Physical Society. (b-c) Reproduced with permission [66]. Copyright 2018, Nature Publishing Group.

**ACKNOWLEDGEMENT**

Y.L. was sponsored in part by Singapore Ministry of Education (No. 218 MOE2018-T2-2-189 (S)), A*Star AME IRG grant (No. A20E5c0095) and Programmatic Funds (No. 219 A18A7b0058), National Research Foundation Singapore Competitive Research Program (No. NRF220 CRP22-2019-0006 and NRF-CRP23-2019-0007). X.L. was sponsored in part by the Fundamental Research Funds for the Central Universities (2021FZZX001-19) and Zhejiang University Global Partnership Fund.

**CONFLIC OF INTERST**

The authors declare no conflict of interest.

**References**:

[1]     Čerenkov, P., "Visible light from pure liquids under the impact of γ-rays," *Dokl. Akad. Nauk SSSR,* Vol. 2, 451-457, 1934.
[2]     Tamm, I. and Frank, I., "Coherent radiation of fast electrons in a medium," *Dokl. Akad. Nauk SSSR,* Vol. 14, 107-112, 1937.
[3]     Ypsilantis, T. and Séguinot, J., "Theory of ring imaging Cherenkov counters," *Nucl. Instrum. Meth. A,* Vol. 343, 30-51, 1994.
[4]     Abashian, A., Gotow, K., Morgan, N., Piilonen, L., Schrenk, S., Abe, K., Adachi, I., Alexander, J., Aoki, K., and Behari, S., "The belle detector," *Nucl. Instrum. Meth. A,* Vol. 479, 117-232, 2002.
[5]     Adam, I., Aleksan, R., Amerman, L., Antokhin, E., Aston, D., Bailly, P., Beigbeder, C., Benkebil, M., Besson, P., and Bonneaud, G., "The DIRC particle identification system for the BaBar experiment," *Nucl. Instrum. Meth. A,* Vol. 538, 281-357, 2005.
[6]     Elder, F., Gurewitsch, A., Langmuir, R., and Pollock, H., "Radiation from electrons in a synchrotron," *Phys. Rev.,* Vol. 71, 829, 1947.
[7]     Ginzburg, V., "Transition radiation and transition scattering," *Phys. Scr.,* Vol. 1982, 182, 1982.
[8]     Happek, U., Sievers, A., and Blum, E., "Observation of coherent transition radiation," *Phys. Rev. Lett.,* Vol. 67, 2962, 1991.

[9]     Smith, S. J. and Purcell, E., "Visible light from localized surface charges moving across a grating," *Phys. Rev.,* Vol. 92, 1069, 1953.
[10]    Koch, H. and Motz, J., "Bremsstrahlung cross-section formulas and related data," *Rev. Mod. Phys.,* Vol. 31, 920, 1959.
[11]    Tian, H., Tice, J., Fei, R., Tran, V., Yan, X., Yang, L., and Wang, H., "Low-symmetry two-dimensional materials for electronic and photonic applications," *Nano Today,* Vol. 11, 763-777, 2016.
[12]    Chen, H., Chan, C. T., and Sheng, P., "Transformation optics and metamaterials," *Nat. Mater.,* Vol. 9, 387-396, 2010.
[13]    Joannopoulos, J. D., Meade, R., and Winn, J. N., *Photonic crystals*. 1995.
[14]    Khanikaev, A. B., Mousavi, S. H., Tse, W.-K., Kargarian, M., MacDonald, A. H., and Shvets, G., "Photonic topological insulators," *Nat. Mater.,* Vol. 12, 233-239, 2013.
[15]    Liu, S., Zhang, P., Liu, W., Gong, S., Zhong, R., Zhang, Y., and Hu, M., "Surface polariton Cherenkov light radiation source," *Phys. Rev. Lett.,* Vol. 109, 153902, 2012.
[16]    Tao, J., Wang, Q. J., Zhang, J., and Luo, Y., "Reverse surface-polariton Cherenkov radiation," *Sci. Rep.,* Vol. 6, 1-8, 2016.
[17]    Massuda, A., Roques-Carmes, C., Yang, Y., Kooi, S. E., Yang, Y., Murdia, C., Berggren, K. K., Kaminer, I., and Soljačić, M., "Smith–Purcell radiation from low-energy electrons," *ACS Photonics,* Vol. 5, 3513-3518, 2018.
[18]    Su, Z., Cheng, F., Li, L., and Liu, Y., "Complete control of Smith-Purcell radiation by graphene metasurfaces," *ACS Photonics,* Vol. 6, 1947-1954, 2019.
[19]    Su, Z., Xiong, B., Xu, Y., Cai, Z., Yin, J., Peng, R., and Liu, Y., "Manipulating Cherenkov radiation and smith–purcell radiation by artificial structures," *Adv. Opt. Mater.,* Vol. 7, 1801666, 2019.
[20]    Hu, H., Gao, D., Lin, X., Hou, S., Zhang, B., Wang, Q. J., and Luo, Y., "Directing Cherenkov photons with spatial nonlocality," *Nanophotonics,* Vol. 9, 3435-3442, 2020.
[21]    Vorobev, V. V. and Tyukhtin, A. V., "Nondivergent Cherenkov radiation in a wire metamaterial," *Phys. Rev. Lett.,* Vol. 108, 184801, 2012.
[22]    Liu, F., Xiao, L., Ye, Y., Wang, M., Cui, K., Feng, X., Zhang, W., and Huang, Y., "Integrated Cherenkov radiation emitter eliminating the electron velocity threshold," *Nat. Photonics,* Vol. 11, 289-292, 2017.
[23]    Chen, H. and Chen, M., "Flipping photons backward: reversed Cherenkov radiation," *Mater. Today,* Vol. 14, 34-41, 2011.
[24]    Rivera, N. and Kaminer, I., "Light–matter interactions with photonic quasiparticles," *Nat. Rev. Phys.,* Vol. 2, 538-561, 2020.
[25]    García de Abajo, F. J. and Di Giulio, V., "Optical excitations with electron beams: Challenges and opportunities," *ACS Photonics,* Vol. 8, 945-974, 2021.
[26]    Xia, F., Wang, H., Xiao, D., Dubey, M., and Ramasubramaniam, A., "Two-dimensional material nanophotonics," *Nat. Photonics,* Vol. 8, 899-907, 2014.
[27]    Koppens, F. H., Chang, D. E., and García de Abajo, F. J., "Graphene plasmonics: a platform for strong light–matter interactions," *Nano Lett.,* Vol. 11, 3370-3377, 2011.
[28]    Lin, X., Yang, Y., Rivera, N., López, J. J., Shen, Y., Kaminer, I., Chen, H., Zhang, B., Joannopoulos, J. D., and Soljačić, M., "All-angle negative refraction of highly squeezed plasmon and phonon polaritons in graphene–boron nitride heterostructures," *Proc. Natl. Acad. Sci,* Vol. 114, 6717-6721, 2017.
[29]    Jiang, Y., Lin, X., and Chen, H., "Directional Polaritonic Excitation of Circular, Huygens and Janus Dipoles in Graphene-Hexagonal Boron Nitride Heterostructures," *Prog. Electromagn. Res.,* Vol. 170, 169-176, 2021.
[30]    Wang, C., Qian, C., Hu, H., Shen, L., Wang, Z., Wang, H., Xu, Z., Zhang, B., Chen, H., and Lin, X., "Superscattering of light in refractive-index near-zero environments," *Prog. Electromagn. Res.,* Vol. 168, 15-23, 2020.

[31] Garcia de Abajo, F. J., "Graphene plasmonics: challenges and opportunities," *Acs Photonics,* Vol. 1, 135-152, 2014.
[32] Zhao, T., Hu, M., Zhong, R., Gong, S., Zhang, C., and Liu, S., "Cherenkov terahertz radiation from graphene surface plasmon polaritons excited by an electron beam," *Applied Physics Letters,* Vol. 110, 231102, 2017.
[33] Tao, J., Wu, L., and Zheng, G., "Graphene surface-polariton in-plane Cherenkov radiation," *Carbon,* Vol. 133, 249-253, 2018.
[34] Rosolen, G., Wong, L. J., Rivera, N., Maes, B., Soljačić, M., and Kaminer, I., "Metasurface-based multi-harmonic free-electron light source," *Light: Sci. Appl.,* Vol. 7, 1-12, 2018.
[35] Pizzi, A., Rosolen, G., Wong, L. J., Ischebeck, R., Soljačić, M., Feurer, T., and Kaminer, I., "Graphene Metamaterials for Intense, Tunable, and Compact Extreme Ultraviolet and X-Ray Sources," *Adv. Sci.,* Vol. 7, 1901609, 2020.
[36] Zhang, X., Hu, M., Zhang, Z., Wang, Y., Zhang, T., Xu, X., Zhao, T., Wu, Z., Zhong, R., and Liu, D., "High-efficiency threshold-less Cherenkov radiation generation by a graphene hyperbolic grating in the terahertz band," *Carbon,* Vol. 183, 225-231, 2021.
[37] Lin, X., Kaminer, I., Shi, X., Gao, F., Yang, Z., Gao, Z., Buljan, H., Joannopoulos, J. D., Soljačić, M., and Chen, H., "Splashing transients of 2D plasmons launched by swift electrons," *Sci. Adv.,* Vol. 3, e1601192, 2017.
[38] Chen, J., Chen, H., and Lin, X., "Photonic and plasmonic transition radiation from graphene," *Journal of Optics,* Vol. 23, 034001, 2021.
[39] Cox, J. D. and Garcia de Abajo, F. J., "Nonlinear interactions between free electrons and nanographenes," *Nano Lett.,* Vol. 20, 4792-4800, 2020.
[40] Kaminer, I., Katan, Y. T., Buljan, H., Shen, Y., Ilic, O., López, J. J., Wong, L. J., Joannopoulos, J. D., and Soljačić, M., "Efficient plasmonic emission by the quantum Čerenkov effect from hot carriers in graphene," *Nat. Commun.,* Vol. 7, 1-9, 2016.
[41] Wong, L. J., Kaminer, I., Ilic, O., Joannopoulos, J. D., and Soljačić, M., "Towards graphene plasmon-based free-electron infrared to X-ray sources," *Nat. Photonics,* Vol. 10, 46-52, 2016.
[42] Geim, A. K. and Grigorieva, I. V., "Van der Waals heterostructures," *Nature,* Vol. 499, 419-425, 2013.
[43] Govyadinov, A. A., Konečná, A., Chuvilin, A., Vélez, S., Dolado, I., Nikitin, A. Y., Lopatin, S., Casanova, F., Hueso, L. E., and Aizpurua, J., "Probing low-energy hyperbolic polaritons in van der Waals crystals with an electron microscope," *Nat. Commun.,* Vol. 8, 1-10, 2017.
[44] Maciel-Escudero, C., Konečná, A., Hillenbrand, R., and Aizpurua, J., "Probing and steering bulk and surface phonon polaritons in uniaxial materials using fast electrons: hexagonal boron nitride," *Phys. Rev. B,* Vol. 102, 115431, 2020.
[45] Tao, J., Wu, L., Zheng, G., and Yu, S., "Cherenkov polaritonic radiation in a natural hyperbolic material," *Carbon,* Vol. 150, 136-141, 2019.
[46] Zhang, Y., Hu, C., Lyu, B., Li, H., Ying, Z., Wang, L., Deng, A., Luo, X., Gao, Q., and Chen, J., "Tunable Cherenkov radiation of phonon Polaritons in silver nanowire/hexagonal boron nitride heterostructures," *Nano Lett.,* Vol. 20, 2770-2777, 2020.
[47] Xi, S., Chen, H., Jiang, T., Ran, L., Huangfu, J., Wu, B.-I., Kong, J. A., and Chen, M., "Experimental verification of reversed Cherenkov radiation in left-handed metamaterial," *Phys. Rev. Lett.,* Vol. 103, 194801, 2009.
[48] Duan, Z., Tang, X., Wang, Z., Zhang, Y., Chen, X., Chen, M., and Gong, Y., "Observation of the reversed Cherenkov radiation," *Nat. Commun.,* Vol. 8, 1-7, 2017.
[49] Duan, Z., Wu, B.-I., Xi, S., Chen, H., and Chen, M., "Research progress in reversed Cherenkov radiation in double-negative metamaterials," *Prog. Electromagn. Res.,* Vol. 90, 75-87, 2009.
[50] Zhang, B. and Wu, B.-I., "Electromagnetic detection of a perfect invisibility cloak," *Phys. Rev. Lett.,* Vol. 103, 243901, 2009.
[51] Hu, H., Lin, X., Zhang, J., Liu, D., Genevet, P., Zhang, B., and Luo, Y., "Nonlocality induced Cherenkov threshold," *Laser Photonics Rev.,* Vol. 14, 2000149, 2020.

[52] Luo, C., Ibanescu, M., Johnson, S. G., and Joannopoulos, J., "Cerenkov radiation in photonic crystals," *Science,* Vol. 299, 368-371, 2003.
[53] Adamo, G., MacDonald, K. F., Fu, Y., Wang, C., Tsai, D., De Abajo, F. G., and Zheludev, N., "Light well: a tunable free-electron light source on a chip," *Phys. Rev. Lett.,* Vol. 103, 113901, 2009.
[54] Song, Y., Jiang, N., Liu, L., Hu, X., and Zi, J., "Cherenkov radiation from photonic bound states in the continuum: towards compact free-electron lasers," *Phys. Rev. Appl.,* Vol. 10, 064026, 2018.
[55] Yang, Y., Massuda, A., Roques-Carmes, C., Kooi, S. E., Christensen, T., Johnson, S. G., Joannopoulos, J. D., Miller, O. D., Kaminer, I., and Soljačić, M., "Maximal spontaneous photon emission and energy loss from free electrons," *Nat. Phys.,* Vol. 14, 894-899, 2018.
[56] Yu, Y., Lai, K., Shao, J., Power, J., Conde, M., Liu, W., Doran, S., Jing, C., Wisniewski, E., and Shvets, G., "Transition Radiation in Photonic Topological Crystals: Quasiresonant Excitation of Robust Edge States by a Moving Charge," *Phys. Rev. Lett.,* Vol. 123, 057402, 2019.
[57] Piazza, L., Lummen, T., Quinonez, E., Murooka, Y., Reed, B., Barwick, B., and Carbone, F., "Simultaneous observation of the quantization and the interference pattern of a plasmonic near-field," *Nat. Commun.,* Vol. 6, 1-7, 2015.
[58] Kurman, Y., Dahan, R., Sheinfux, H. H., Wang, K., Yannai, M., Adiv, Y., Reinhardt, O., Tizei, L. H., Woo, S. Y., and Li, J., "Spatiotemporal imaging of 2D polariton wave packet dynamics using free electrons," *Science,* Vol. 372, 1181-1186, 2021.
[59] Wang, K., Dahan, R., Shentcis, M., Kauffmann, Y., Hayun, A. B., Reinhardt, O., Tsesses, S., and Kaminer, I., "Coherent interaction between free electrons and a photonic cavity," *Nature,* Vol. 582, 50-54, 2020.
[60] Dahan, R., Nehemia, S., Shentcis, M., Reinhardt, O., Adiv, Y., Shi, X., Be'er, O., Lynch, M. H., Kurman, Y., and Wang, K., "Resonant phase-matching between a light wave and a free-electron wavefunction," *Nat. Phys.,* Vol. 16, 1123-1131, 2020.
[61] Vanacore, G. M., Berruto, G., Madan, I., Pomarico, E., Biagioni, P., Lamb, R., McGrouther, D., Reinhardt, O., Kaminer, I., and Barwick, B., "Ultrafast generation and control of an electron vortex beam via chiral plasmonic near fields," *Nat. Mater.,* Vol. 18, 573-579, 2019.
[62] Galiffi, E., Huidobro, P., and Pendry, J., "Broadband nonreciprocal amplification in luminal metamaterials," *Phys. Rev. Lett.,* Vol. 123, 206101, 2019.
[63] Oue, D., Ding, K., and Pendry, J., "Cerenkov radiation in vacuum from a superluminal grating," *arXiv preprint* Vol. arXiv:.13681, 2021.
[64] Sloan, J., Rivera, N., Joannopoulos, J. D., and Soljačić, M., "Two photon emission from superluminal and accelerating index perturbations," *arXiv preprint* Vol. arXiv:.09955, 2021.
[65] Ginis, V., Danckaert, J., Veretennicoff, I., and Tassin, P., "Controlling Cherenkov radiation with transformation-optical metamaterials," *Phys. Rev. Lett.,* Vol. 113, 167402, 2014.
[66] Lin, X., Easo, S., Shen, Y., Chen, H., Zhang, B., Joannopoulos, J. D., Soljačić, M., and Kaminer, I., "Controlling Cherenkov angles with resonance transition radiation," *Nat. Phys.,* Vol. 14, 816-821, 2018.
[67] Lin, X., Hu, H., Easo, S., Yang, Y., Shen, Y., Yin, K., Blago, M. P., Kaminer, I., Zhang, B., Chen, H., Joannopoulos, J., Soljačić, M., and Luo, Y., "A Brewster route to Cherenkov detectors," *arXiv preprint,* Vol. arXiv:.11996, 2021.
[68] Hu, H., Lin, X., Wong, L. J., Yang, Q., Zhang, B., and Luo, Y., "Surface Dyakonov-Cherenkov Radiation," *arXiv preprint* Vol. arXiv:.09533, 2020.